# Nutrient Loading Increases Red Snapper Production in the Gulf of Mexico

Joshua M. Courtney,[a] Amy C. Courtney,[a] and Michael W. Courtney[b]

a BTG Research, P.O. Box 62541, Colorado Springs, CO, 80962
b United States Air Force Academy,[1] 2354 Fairchild Drive, USAF Academy, CO, 80840

**Abstract**
A large, annually recurring region of hypoxia in the northern Gulf of Mexico has been attributed to water stratification and nutrient loading of nitrogen and phosphorus delivered by the Mississippi and Atchafalaya rivers.  This nutrient loading increased nearly 300% since 1950, primarily due to increased use of agricultural fertilizers.  Over this same time period, the red snapper (*Lutjanus campechanus*) population in the Gulf of Mexico has shifted strongly from being dominated by the eastern Gulf of Mexico to being dominated by the northern and western Gulf of Mexico, with the bulk of the current population in the same regions with significant nutrient loading from the Mississippi and Atchafalaya rivers and in or near areas with development of mid-summer hypoxic zones.  The population decline of red snapper in the eastern Gulf is almost certainly attributable to overfishing, but the cause of the population increase in the northern and western Gulf is subject to broad debate, with the impact of artificial reefs (primarily oil platforms which have increased greatly since the 1960s) being the most contentious point.  Nutrient loading has been shown to positively impact secondary production of fish in many marine systems.  The present paper offers the hypothesis that increased nutrient loading has contributed significantly to increased red snapper population in the northern and western Gulf of Mexico.  Nutrient loading may be working in synergy with the abundant oil platforms both increasing primary production and providing structure encouraging red snapper to feed throughout the water column.

**Corresponding Author**
Michael W. Courtney  ·United States Air Force Academy, 2354 Fairchild Drive, USAF Academy, CO, 80840 Email: Michael_Courtney@alum.mit.edu

**Introduction**
The authors have been following the literature on nutrient loading and hypoxia in the Gulf of Mexico [1-3] for several years.  Characterizations of these areas as "dead zones" perplexed us, because we had personal experience with the Louisiana-Texas shelf teeming with life, and we knew that Louisiana harvests tremendous quantities of seafood and also that Louisiana supports one of the best sport fisheries in the Gulf of Mexico.  Our anecdotal observations and review of available data seemed to suggest that several sport species of fish (red drum, spotted sea trout, red snapper) are both plumper and more plentiful in Louisiana waters, and we have often considered hypothetical explanations for these observations.

At the same time, we became aware of the attraction vs. production debate regarding whether oil platforms and other artificial reefs only attract red snapper for more efficient harvest or whether they make significant contributions to increased production of red snapper in the Gulf of Mexico [4-6]. While reviewing Melissa Monk's doctoral thesis [7] (Monk 2012, Figures 1.14 – 1.22),  the apparent decrease in biomass with increasing distance from the Mississippi River discharge as well as the apparent greater biomass in the areas with higher nutrient loading and lower oxygen levels caught our attention. Monk's suggestion that Mississippi River outflow and nutrient loading should be considered

---

to explain the temporal variability in both total biomass and how the biomass is distributed among species also stood out to us.

Consideration of these ideas led us to the hypothesis that nutrient loading from the Mississippi River makes a significant contribution to the increases in the red snapper population that have been observed in the northern and western Gulf of Mexico. Since we were also aware of the artificial reef hypothesis, it took more time and consideration of a broad array of literature concerning red snapper life history to weigh whether the artificial reefs were merely coincidental before refining the hypothesis to include synergies by which both artificial reefs and the nutrient loading contribute to red snapper production. The main benefits of nutrient loading are probably the increase in available forage for red snapper and perhaps relieving predatory pressure on juvenile fish. Benefits of artificial reefs include providing habitat and protection from predation for younger fish and also allowing red snapper to feed throughout the water column rather than only on the bottom.

Red snapper life history includes a reproductive strategy of early reproduction (age 2-4) and producing many millions of eggs over a life span. Very few eggs survive to adulthood [5]. The majority of eggs are laid in 50-100 m of water and drift to shallower water before hatching. Eggs and larvae remain pelagic (drifting freely in open water neither close to the bottom nor the shore) until metamorphosis and settlement at close to 27 days and 18 mm total length. Most pre-recruit red snapper either settle on or quickly move to structured habitat such as low relief shell habitat, artificial reefs, or structured, hard bottom in 20-50 m depths. The age zero fish grow quickly and seek larger, more structured habitat to maintain protection from predation. Age 1 fish move to larger, more complex reefs in winter, allowing age 0 red snapper to move into the lower relief habitats they are abandoning. The Gulf of Mexico is dominated by soft bottom with very little natural reef habitat with vertical relief over 1 m. By age 2, many red snapper occupy artificial reefs (those providing > 1 m of vertical relief) including petroleum platforms, and many have moved to deeper water, 50-100 m. Red snapper show high site fidelity ($\sim 0.5$ yr$^{-1}$ is typical), but can be displaced significant distances by tropical storms. When displacement occurs, the slow migration tends to be more eastward (toward Florida) and southward (toward deeper water and more shelf-edge habitat) as red snapper age from 2 to 8. Fish over 8 years are not as susceptible to predation and are often found over soft, open areas of the sea floor.

In support of the hypothesis that nutrient loading increases red snapper production, this paper first summarizes nutrient loading in the Gulf of Mexico, reviews how nutrient loading has been shown to increase secondary production in other marine systems, then describes the decline of the red snapper fishery (due to overfishing) and the debate surrounding alternate explanations for the increase in red snapper production in the area since it reached a minimum in the late 1980s. The hypothesis of the importance of nutrient loading is then explained along with the possible synergies of how nutrient loading and artificial reefs work together to increase red snapper production. Finally, the testability of the hypothesis is discussed along with possible applicability in other ecosystems. Considering the supporting background, it is hard to believe that this hypothesis has not been raised before, and the authors are reluctant to claim novelty. However, the hypothesis of nutrient loading does not play a major role in the red snapper production debate, nor is the increase in secondary production of any species of finfish (much less red snapper) a prominent feature in published papers centered on eutrophication and hypoxia in the Gulf of Mexico.

Grimes [8] pointed out that the fact that 70-80% of Gulf of Mexico fishery landings come from waters surrounding the Mississippi delta provided strong circumstantial evidence that nutrient enhancement enhances fishery production, and this is consistent with a large body of work suggesting riverine discharge nutrient enhances fisheries [9]. Grimes [8] suggested that two major groupings, estuary dependent species such as red drum (*Sciaenops ocellatus*), spot (*Leiostomus xanthurus*), and Atlantic croaker (*Micropogonias undulatus*), and coastal species such as king mackerel (*Scomberomorus* *calvalla*), Spanish mackerel (*Scomberomorus maculatus*), and bluefish (*Pomatomus saltatrix*) were most likely to be influenced by this nutrient enrichment. Grimes pointed out that while nutrient enhancement of fishery production seemed clear, the exact mechanisms remained unclear, and Grimes did not suggest enhancement for reef dependent demersal fishes such as the red snapper.

**Nutrient Loading and Hypoxia in the Gulf of Mexico**
Nutrient loads of nitrogen and phosphorus being delivered to the Gulf of Mexico by the Mississippi and Atchafalaya rivers increased markedly in the second half of the 20$^{th}$ century [1,2]. The mean annual concentration of nitrate was comparable in 1905-1906 and 1933-1934 when compared with the 1950s. However, the mean annual nitrate concentration tripled from 1955 to 1996 with most of the increase occurring between 1970 and 1983. This increase in nitrate concentration has some contribution from point sources but is dominated by the increased use of agricultural fertilizers throughout the Mississippi River watershed. The records for phosphorus loading before 1973 are not as reliable. Estimates suggest phosphorus loads have approximately doubled over the same time period, though these estimates are largely based on extrapolation of a regression into the past rather than direct measurements.

There is broad agreement that this nutrient loading is the principal cause of large hypoxic zones that occur annually on the Louisiana-Texas shelf in the Gulf of Mexico. The Louisiana-Texas shelf extends westward from the mouth of the Mississippi River (near 271° W longitude) to the eastern coast of Texas (near 263° W longitude) and southward from the Louisiana-Texas coast (near 29.5° N latitude) to the 200 m depth contour (near 27.5° N latitude). Maps of the areas impacted by hypoxia are available in many references [1-3] and web based maps for each year from 2001 to 2012 are available online (http://www.ncddc.noaa.gov/hypoxia/products/). Areas of oxygen depleted water at or near the bottom begin to form after the nutrient flux peaks (usually in May), tend to reach a maximum size in July or August, and then dissipate in the fall when water temperatures cool and wind associated with passing cold fronts causes mixing of well oxygenated surface waters and oxygen depleted bottom layers. Mixing caused by tropical storms during the summer can also reduce the size of hypoxic zones [1-3].

In addition to nutrient loading, stratification of the water column on the Louisiana-Texas shelf also contributes to the formation and maintenance of the hypoxic zones. Lower salinity waters from the Mississippi and Atchafalaya rivers are less dense than higher salinity lower layers. Waters higher in the water column also tend to experience more warming which maintains the stratification (warmer water is less dense). The nutrient loading hastens and facilitates the growth of algae and phytoplankton, which, in turn, can also cause a rapid increase in the biomass of zooplankton and nekton. When the phytoplankton and algae die, they sink to the bottom. Likewise, the increased biomass of zooplankton and nekton produce an abundance of feces which also sinks to the bottom. The breakdown of the dead algae, phytoplankton, and feces by aerobic bacteria near the bottom deplete the oxygen near the bottom much faster than it can be replenished by the low rates of mixing with the well-oxygenated surface waters [1-3].

There are many nuances to hypoxia formation, including contributions from carbon loading from Mississippi River water [3], and the nutrient loading and occasional formation of hypoxic zones in the northern Gulf of Mexico off the coasts of Mississippi and Alabama in years when currents move substantial nutrients eastward from the Mississsippi River discharge [2]. Variations in nutrient loading, fresh water flow from the rivers, and stratification can cause wide variations in the resulting area of hypoxic water. For example, since the regular mapping of the hypoxic zone began in 1985, the hypoxic area has varied from under 5000 km$^2$ in 1988 and 2000 to near 20,000 km$^2$ in 1999, 2001, 2002, and 2007 [2].

**Nutrient Enrichment and Increased Secondary Production of Marine Systems**
Nixon and Buckley [10] review evidence that secondary production in marine systems tends to be increased by nutrient enrichment. One example is the collapse of the fishery on the Egyptian shelf after the closing of the Aswan dam in 1965. This restricted the deposition of inorganic nutrients by the annual Nile flood; subsequently the fishery collapsed. However, use of inorganic fertilizers and point source discharge of nutrients increased markedly in the Nile delta after that time, and the resulting increase in inorganic nitrogen coincided with a dramatic recovery of the fishery that began in the early to mid 1980s and has continued to the present. Caddy [11] also discusses the impact of the Aswan dam on the fisheries off the Nile delta and the later rise in production with enriched drainage water in the early to mid 1980s.

A second example discussed by Nixon and Buckley [10] are Scottish sea loch experiments with fertilization. Though the experiments showed little promise regarding the potential commercial success of estuary fertilization due to loss of fish to hypoxia and escape into the open ocean, the experiments demonstrated remarkable increases in benthic fauna (5-6 times in one of the sea lochs) in spite of heavy feeding from introduced fish. These experiments demonstrated remarkable growth rates of plaice (*Pleuronectes platessa*).

A third example discussed by Nixon and Buckley [10] is the Baltic Sea, which has received increasing amounts of nitrogen and phosphorus from agricultural runoff, urban wastewater, and enriched atmospheric deposition for many decades. Between an early study in 1920-1923 and a later assessment in 1976-1977, the benthic biomass above the halocline was estimated to increase by 2-10 times. Marked decreases in benthic biomass below the halocline over the same interval were attributed to decreases in oxygen due due to eutrophication. Nixon and Buckley also describe Thurow's historical reconstruction [12] of biomass and finfish yield in the Baltic Sea showing that both biomass and finfish yield were low in the first half of the 20th century, then increased sharply after about 1950. After considering a number of other possible factors [12], Thurow concluded that nutrient loading and subsequent eutrophication were responsible for the 8-fold increase in fish biomass. In response to some who argued for reduction of nutrient inputs, Thurow [12] warned that reduction of nitrogen and phosphorus inputs would "inevitably lead to a lower fish biomass."

Caddy [13] described a simple conceptual model relating fishery landings to nutrient inputs. There is a positive correlation up to some point, above which higher nutrient loadings would cause fishery production to decrease. Caddy speculated that at higher levels of nutrient loadings (above a system dependent threshold) additional nutrients would cause seasonal or permanent bottom anoxia leading to a decline in benthos (bottom dwelling) and benthos-feeding fish. In addition to the examples of production increasing with nutrient load reviewed by Nixon and Buckley [10], Caddy [11] describes the increases in production in the northern Mediterranean with increased nutrient load, although some areas (the northern Adriatic) have experienced some years of hypoxia. Oczkowski and Nixon [14] point out that even though the positive response of fishery production to nutrients has been demonstrated in many cases [10], (Nixon and Buckley 2002) the dome shaped response predicted by Caddy has not been widely demonstrated in the literature. In a detailed review of available data for nitrogen loadings and fishery production in the Nile delta of Egypt, Oczkowski and Nixon [12] showed fish landings increased with nitrogen loading up to a threshold concentration of approximately 100 µM of dissolved inorganic nitrogen; above this threshold, there was an exponential decrease in landings. Breitberg [15] also pointed out that high productivity of finfish in nutrient-enriched systems are often greater than losses due to oxygen depletion.

**Red Snapper in the Gulf of Mexico**
Commercial fishing for red snapper in the Gulf began on a significant scale in the late 1800s. Until 1950, the red snapper harvest in the eastern Gulf dominated the harvest. Exploratory trips were made to the western Gulf, including the Louisiana-Texas shelf, but this area was found to be relatively unproductive. From 1950 until the present, the red snapper harvest in the western Gulf has demonstrated an increasing trend, while the production and harvest of the eastern Gulf has steadily declined [4, 5].

In addition to a shift in production from the eastern to the western Gulf, estimates of the sustainable yield in the Gulf have also increased markedly. The highest annual harvests before 1970 were approximately 9 million pounds (mp) and in most years were about 5 mp [4]. These yields were not sustainable and led to a collapse of the fishery in the eastern Gulf between 1980 and the early 1990s. However, current model projections of maximum sustainable yield for the Gulf are between 11.3 and 25.4 mp annually, with most of that production dominated by the northern and western Gulf, especially the Louisiana-Texas shelf [4-6].

There is broad agreement that the collapse of the red snapper stocks in the Gulf was due to overfishing [4-6, 16]. In contrast, there has been lively debate regarding the causes most strongly related to the stock recovery in the western Gulf (the Louisiana-Texas shelf) and in the northern Gulf (off the Alabama coast). The hypothesis that artificial reefs have played an important role finds strong support in the empirical evidence [4, 5, 17]. Red snapper in this area are limited by available habitat, which is dominated by sand, silt, and mud bottoms with limited hard bottom or reef which play a key role at multiple stages of red snapper life history. In 1960, there were only about 351 offshore petroleum platforms in the western and northern Gulf. The number increased to 1520 platforms by 1970 and since 1990 has remained steady at around 4000. Petroleum platforms provide valuable three dimensional artificial reef habitat by extending from the bottom to the surface. Because snapper segregate by depth at petroleum platforms (typically younger fish at shallower depths and older fish deeper), a given area of petroleum platform provides much more habitat than the relatively scarce low relief (most < 1m high) natural reefs in the area. Petroleum platforms support fish densities 10 to 1000 times greater than adjacent sand and mud bottoms and also exceed densities supported by natural hard bottom. In addition, from 1950 to 2009, an estimated 20,000 artificial reef structures have been placed in a 1200 square mile area off the Alabama coast. Though lower relief (< 2m typically) than oil platforms, these reefs were deployed in areas dominated by sandy mud bottoms, having limited natural relief (a few cm) and dominated by diminutive fish species [4, 5]. Figure 4 of Shipp and Bartone [4] shows a map of the oil platforms in the northern Gulf of Mexico, and an interactive data map which facilitates overlaying the petroleum platform locations with depth contours and a variety of physical and biological parameters is available online (http://www.ncddc.noaa.gov/website/DataAtlas/atlas.htm#).

The debate often centers around whether these artifical reefs only attract fish or whether they contribute significantly to production [6]. The attraction viewpoint is based strongly on the assumption that red snapper do not recruit heavily to artificial reefs until age 2, which in turn suggests the artificial reefs are unlikely to relieve a life history bottleneck [6]. In contrast, there is ample evidence that red snapper begin using artificial reefs much earlier in their life history [4, 5, 18]. The attraction viewpoint has also failed to provide compelling evidence for where the increased production of red snapper is coming from if artificial reefs are not contributing significantly. In contrast, additional support for the production hypothesis is found in correlations of red snapper age with the age of the artificial reef [19], greater utilization by red snapper of artificial reefs with epibenthic communities [20], higher site fidelity [21] than was presumed by Cowan et al. [6], and density dependent mortality of age 0 fish [22].

**Hypothesis**
Authors favoring the attraction-only viewpoint of artificial reefs in the Gulf have asserted that high densities of benthic feeding predators will eventually create a feeding halo around the reef and lead to necessary longer foraging distances over soft mud bottoms [23-25]. In addition, some authors have viewed the lack of decline in the shallow Gulf shelf biomass as a paradox in light of the seemingly increasing area of the hypoxic zone [1]. The possibility of the Gulf shelf being on the increasing part of the Caddy curve may not have been considered, even while relating a 4 fold increase in biomass between 2002 and 2010 and citing Caddy [13]. It seems ironic that the areas where biomass has increased 4 fold are portrayed as "dead zones."

The new hypothesis we offer is that the nutrient loading in the northern and western Gulf of Mexico contributes significantly to the observed increases in red snapper production. The nutrient loading probably has multiple synergistic effects with the vertical reef structures (petroleum platforms), and by increasing biomass and productivity in forage species (red snapper eat almost any available forage) probably put enough forage within relatively short forage distances of most artificial reefs that the halo effect (if it exists at all) is of much less importance than in less productive ecosystems. In a recent study, Daigle [26] found no halo effect around Louisiana shelf oil platforms that were well populated with red snapper, and no decrease of common red snapper prey items in the areas adjacent to the oil platforms.

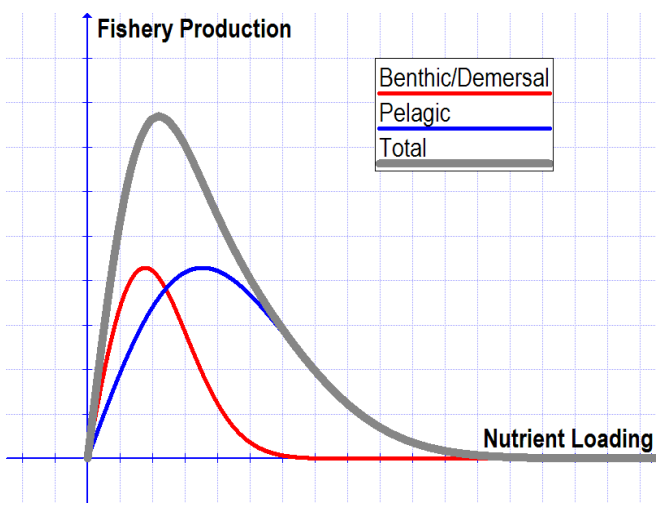

*Figure 1: Caddy curve [11, 13] showing that fishery production likely peaks with some optimal amount of nutrient loading and then sharply declines with higher loading due to eutrophication and hypoxia. Because hypoxia tends to impact bottom waters more strongly, the figure shows conceptual curves for benthic and demersal production separately from pelagic production which is likely to respond positively to greater nutrient loads before a sharp decrease as hypoxia begins to negatively impact the entire water column.*

Figure 1 shows the conceptual Caddy curve [11, 13]. The basic ideas are intuitive and obvious, though the details and thresholds will depend on the specific ecosystem. Oczkowski and Nixon [10] found that Egyptian Nile delta productivity increased at nitrogen concentrations up to 100 µM and then declined. Mississippi River nitrogen concentrations tend to be about 100 µM before dilution with Gulf of Mexico salt water. Given the estimate of a four fold increase in overall biomass on the Louisiana-Texas shelf [1] and the increase in productivity of red snapper coincident with the increase in nutrient loading, red snapper production may be on the rising edge or near the top of the response curve of the species to nutrient loading.

In addition to providing hard substrate for many species near the bottom of the water column, petroleum platforms provide both hard substrate and cover throughout the water column. Hard substrate likely increases primary production and allows a variety of reef dwelling organisms to live throughout the water column. Structure provides cover from predators. This enables red snapper to occupy a pelagic role in the food web (feeding throughout the water column), in addition to their more commonly accepted demersal role as a reef fish (feeding near the bottom). McCawley [24] found that even around the relatively low relief (< 2m) artificial reefs off the Alabama coast, red snapper diet included 41% (by weight) of prey items residing in the water column with 55% (by weight) of benthic prey items. Isotope analysis has suggested that red snapper on oil platforms obtain approximately 55.5% of their nutrition from platform dependent food web contributions and 44.5% of their nutrition from the background food web [26, Table 3.6, p. 88].

The more pelagic role facilitated by petroleum platforms may have the effect of creating a broad, relatively flat near-optimum region in the red snapper production response to increasing nutrient loading as the population is able to shift from benthic to pelagic feeding in years where nutrient loading greatly increases pelagic production or the seasonal hypoxic zone depletes benthic resources. de Leiva Monero et al. [28] found that, averaged across semi-enclosed European seas, pelagic production increased from 0.89 Mt/km$^2$ to 2.30 Mt/km$^2$ between mesotrophic and eutrophic systems; whereas, demersal production was flat, averaging 0.83 Mt/km$^2$ for both mesotrophic and eutrophic systems. It may be fortuitous that most artificial reefs on the Louisiana-Texas shelf are petroleum platforms that extend from bottom to surface, providing structure for red snapper habitat at whatever level of the water column happens to be most productive at a given time; whereas, off the coast of Alabama, most of the artificial reefs are relatively low relief. This area of the shelf benefits from some level of nutrient enrichment from the Mississippi river discharge, but since the dominant currents move most of the Mississippi River plume westward, hypoxia off the coast of Alabama is less common.

In addition to the mechanism of increasing available forage for red snapper resident at artificial reefs, nutrient loading may contribute to red snapper production via other mechanisms:

1. Nutrient loading increases productivity for important forage sources at key life stages.
2. The zone of hypoxia reduces the local population of predators that severely limited production at some life stage bottleneck.
3. At some key life stage, red snapper prey heavily upon forage fleeing from the hypoxic zone or recently deceased forage being moved out of the hypoxic zone by currents.
4. When the hypoxic zone dissipates in the fall, the fauna that quickly populate it provide more abundant forage than the fauna that dominate in the absence of hypoxic events.
5. Nutrient loading increases availability of cover at key life stages where available cover was an important bottleneck.

**Discussion: Testability and Broader Applicability**
The hypothesis that nutrient loading in the northern and western Gulf contributes significantly to red snapper production would be supported by the following:

1. Positive correlation of increased red snapper production (or a given life history stage) with nutrient loading. Since nutrient loading varies considerably spatially and temporally, support will be stronger if the observations are more localized.
2. Positive correlation of increased red snapper body condition (for example, using the relative condition factor, Kn) with nutrient loading. Relative condition factor is more sensitive than overall production or growth rates to seasonally or spatially varying factors.
3. The relative importance of hypoxia vs. nutrient loading should be considered. It is expected that nutrient loading in the absence of hypoxia will benefit red snapper more than cases where the same level of local nutrient loading is accompanied by local hypoxia.
4. Observations of nutrient loading leading to increased (probably vegetative) cover and observations of red snapper making use of vegetative cover at a key life history stage.

The principles and mechanisms that may be involved likely have broader applicability for interpreting and perhaps mitigating other eutrophic systems.  The hypothesis that a valuable commercial fish species may be able to transition from demersal to pelagic to take advantage of (rather than suffer in) a eutrophic environment may have important analogs in other systems which tend to be dominated by small pelagic fishes once seasonal hypoxia begins [11].  The challenge for fisheries managers is to find and determine how to make best use of this possible silver lining in systems strongly impacted by nutrient loading and eutrophication.

Speculation on specific applications of this hypothesis in other ecosystems is tentative at best.  However, there are many other shelf areas in the world affected by nutrient loading or likely to be affected by increasing levels of nutrient loading as food production and population increases in the watershed areas.  Nutrient loading can also decrease in response to economic downturns and efforts at mitigation.  Application to other ecosystems may be less speculative after the mechanistic details are better described and verified for the case of red snapper, especially regarding the role of petroleum platforms.

Available evidence suggests that primary production in the near shore waters surrounding the Yangtze river plume in the East China Sea is not likely nutrient limited due to high turbidity, short residence time, and strong tidal mixing [29].  Levels of chlorophyll a are much lower in the areas around the Yangtze river plume than the plumes of the Mississippi, Amazon, and Danube rivers.  In addition to nutrient load, enhanced production depends on the ability of light to penetrate for photosynthesis as well as weaker mixing so that algae remains in the surface waters for longer periods of time.  The impacts of nutrient loading on fisheries production are strongly confounded with effects of overfishing in this area.  The Amazon river plume region is likely still on the rising edge of the Caddy curve, and fisheries yields of important commercial species of both pelagic and demersal species are likely to increase somewhat in response to increased nutrient loading.  However, there are so many varying features of each food web that the existing knowledge base is likely inadequate for making specific predictions to any given system.  In contrast, the existing knowledge base is more appropriate to suggest possible effects and important areas of observation as nutrient loading changes in response to natural and anthropogenic changes in watersheds.

**Acknowledgments**
This research was funded by BTG Research (www.btgresearch.org) and the United States Air Force Academy.  We thank the reviewers and editors for helpful comments that improved the manuscript.